\documentclass[12pt]{article}

\usepackage{graphicx}
\usepackage{wrapfig}
\usepackage{boxedminipage}
\usepackage{setspace}
\usepackage{subfigure}

\begin{document}

\baselineskip 6mm
\renewcommand{\thefootnote}{\fnsymbol{footnote}}
\newcommand{\nc}{\newcommand}
\newcommand{\rnc}{\renewcommand}


\rnc{\baselinestretch}{1.24}    
\setlength{\jot}{6pt}       
\rnc{\arraystretch}{1.24}   

\makeatletter \rnc{\theequation}{\thesection.\arabic{equation}}
\@addtoreset{equation}{section} \makeatother



\def\be{\begin{equation}}
\def\ee{\end{equation}}
\def\ba{\begin{array}}
\def\ea{\end{array}}
\def\bea{\begin{eqnarray}}
\def\eea{\end{eqnarray}}
\def\nn{\nonumber\\}


\def\ct{\cite}
\def\lb{\label}
\def\eq#1{Eq. (\ref{#1})}


\def\a{\alpha}
\def\b{\beta}
\def\g{\gamma}
\def\G{\Gamma}
\def\d{\delta}
\def\D{\Delta}
\def\ep{\epsilon}
\def\e{\eta}
\def\ph{\phi}
\def\Ph{\Phi}
\def\ps{\psi}
\def\Ps{\Psi}
\def\k{\kappa}
\def\l{\lambda}
\def\L{\Lambda}
\def\m{\mu}
\def\n{\nu}
\def\th{\theta}
\def\Th{\Theta}
\def\r{\rho}
\def\s{\sigma}
\def\S{\Sigma}
\def\ta{\tau}
\def\o{\omega}
\def\O{\Omega}
\def\pr{\prime}


\def\half{\frac{1}{2}}

\def\goto{\rightarrow}

\def\na{\nabla}
\def\grad{\nabla}
\def\curl{\nabla\times}
\def\div{\nabla\cdot}
\def\pa{\partial}

\def\la{\langle}\def\ra{\rangle}
\def\ll{\left\langle}
\def\rr{\right\rangle}
\def\lb{\left[}
\def\lc{\left\{}
\def\ls{\left(}
\def\ld{\left.}
\def\rd{\right.}
\def\rb{\right]}
\def\rc{\right\}}
\def\rs{\right)}

\def\vac#1{\mid #1 \rangle}


\def\check{ \maltese {\bf Check!}}


\def\Tr{{\rm Tr}\,}
\def\det{{\rm det}}


\def\bc#1{\nnindent {\bf $\bullet$ #1} \\ }
\def\ch {$<Check!>$ }
\def\ss {\vspace{1.5cm}}
\begin{titlepage}

\hfill\parbox{5cm} { }

\vspace{25mm}

\begin{center}
{\Large \bf Gluon Condensation at Finite Temperature via AdS/CFT}

\vskip 1. cm
  {\large Youngman Kim$^{a}$, Bum-Hoon Lee$^{b}$,
  Chanyong Park$^{b}$ and  Sang-Jin Sin$^{c}$}
\vskip 0.5cm
{\it $^a\,$School of Physics, Korea Institute for Advanced Study,
Seoul 130-722, Korea}\\
 { \it $^b\,$ CQUeST, Sogang University,
Seoul 121-742, Korea}\\
 {\it  $^c\,$ Department of physics, BK21 Program Division,
Hanyang University, Seoul 133-791, Korea}
\end{center}
\thispagestyle{empty}

\vskip2cm


\centerline{\bf ABSTRACT} \vskip 4mm

\vspace{1cm}
We consider gluon condensation (GC) at finite temperature  using AdS/CFT. We first show
that in the presence of regular horizon, the GC is forbidden in high temperature.
Then we consider gravity back-reaction to dilaton coupling and
show that the back-reaction develops an singularity, and
non-vanishing value of gluon condensation is allowed.
We also study thermodynamic quantities and the trace anomaly in the presence of the GC.
We discuss how to define a temperature in the presence of the
singularity which forbids Hawking temperature.  Finally we
describe the thermodynamics of the gluon condensation
including the effect of the Hawking-Page transition.

\vspace{2cm}


\end{titlepage}

\renewcommand{\thefootnote}{\arabic{footnote}}
\setcounter{footnote}{0}

\section{Introduction}

There are increasing activities in using AdS/CFT
\cite{AdS/CFT} to describe the real systems after relevant
deformations of the AdS background. For example, it has been suggested that the fireball in Relativistic Heavy Ion Collision (RHIC)  should be considered as a
strongly interacting system \cite{RHIC-1,RHIC-2} and the fireball
has been studied using dual gravity models
\cite{son,SZ,Nastase,yaffe,jns,gubser,SZ2,gubser2}.
In the finite temperature context
the SUSY is broken, and therefore the models have more chance to be in the same universality class of real QCD.
Many attempts were made to construct models phenomenologically
closer to QCD \cite{AdS/QCD} as well
as models with  mesons and quenched  quarks \cite{kk,evans,myers1,myers2}.

More recently, in
an interesting  paper \cite{CR},  dilaton-gravity solution   describing the gluon condensation at zero temperature \cite{CR} was discussed.  We call this solution as dilaton-wall solution.
 In fact,  dilaton-wall solution has a rather long history of being  repeatedly  rediscovered \cite{nojiri,sfet,gubser3}.
 The gluon condensate was originally
 introduced, at zero temperature, by Shifman, Vainshtein and Zakharov (SVZ) as a measure for
 nonperturbative physics in QCD~\cite{SVZ}.
 At high temperature, it is useful to
 study the nonperturbative nature of the quark-gluon plasma (QGP), and it can be served as an order parameter for
 (de)confinement~\cite{SHLee, EGM, Miller}. Recently the role of the
 gluon condensate in RHIC physics is extensively studied in ~\cite{BR06}.
Here motivated by the recent paper \cite{CR},
we describe the temperature dependence of the gluon condensation
by extending   dilaton-wall solution to the finite temperature. 
We consider  the   back-reaction of the ads black hole solution and 
the immediate consequences are of two folds:
one is the development of the singularity (instead of regular horizon) and the other is
the breaking of the conformal symmetry.
  We shows that  the singularity  plays a crucial  role to
  allow the non-vanishing value of gluon condensation. 
   We also study thermodynamic quantities and the trace anomaly in the presence of the GC.
Although the singularity of the solution does not admit the Hawking temperature,
we argue that we can nevertheless define the temperature using the holographic
correspondence of an extended Stephan-Boltzmann law.  Finally we  describe the
temperature dependence of the gluon condensation  including the effect of the
Hawking-Page transition (HPT).

The rest of the paper goes as follows. 
In section 2, we first consider the hard wall model~\cite{PS}, and then we discuss the back-reaction due to the gluon condensation in section 3. Our solution interpolates the dilaton-wall solution and
the well-known AdS black hole solution.
Since the  solution does not admit the Hawking temperature in generic parameter values,
 we define the temperature using the temperature in the absence of condensation and justify  using the holographic correspondences. 
In section 4, we will discuss the thermodynamics of the gluon condensation by calculating various thermodynamic potentials.  In section 5, we  consider temperature dependence  of the gluon condensation
 including Hawking-Page  transition~\cite{hawkingpage,witten2,herzog}.
In the discussion section, we describe some  limitations of the present approach and a few future works.

\section{Gluon condensation in hard wall model}
The gluon condensation is in fact the simplest quantity in AdS/CFT
consideration, since it couples with the dilaton which is a massless scalar. Here we use a simple model, called hard wall model,
 where  confinement is treated by introducing the IR cut off.
 The action is given by a scalar coupled with background gravity minimally.
 \be
 S=-\frac{1}{2\kappa^2} \int d^4x dz \sqrt{g} \partial_\mu\phi \partial^\mu \phi .\ee
The equation of motion for this massless field is
 \be
 \pa_z (z^{-3} f(z) \pa_z\phi )=0.
 \ee
In low temperature, the thermal AdS background is dominating for which $f(z)=1$, and the solution is given by
\be
\phi(z)=c_0+c_1z^4.
\ee
 If we have a hard wall at $z_m$ and UV boundary at $z=0$,
the proper boundary should be the Dirichlet  boundary condition (BC)
 at the wall and at the AdS boundary $z=0$:
\be
\phi(z_m)=A, \quad \phi(0)=B .
\ee
These boundary conditions determine $c_i$'s as  $c_0=B$, and
$c_1=(A-B)/z_m^4$.
On the other hand, the general AdS/CFT dictionary identifies $c_1$ as the gluon condensation
\be
c_1\sim <\Tr F_{\mu\nu}^2>.
\ee
The point here is that it is temperature independent in a confining phase, simply because there is no explicit temperature dependence in the background metric.
In fact this is consistent with the recent result in large N gauge theory
result \cite{yaffe2}. \footnote{ The
  temperature independence of  the chiral condensation were
   reported long ago \cite{neri,pisarski}, and
   in the AdS/CFT context we  argued in the same line here \cite{KSJL}.}
The temperature dependence
is suppressed for  chiral condensation as well as gluon condensation in the gauge theory context.

According to Herzog\cite{herzog}, there is a
Hawking-Page type transition at the $T=2^{1/4}/\pi z_m$,
and beyond this temperature the AdS black hole background is dominant.
\be \label{dbh}
ds^2 = \frac{R^2}{z^2} \lb dz^2 +
   \ls  {1+   az^4  }  \rs
   \ls d \vec{x}^2 -
    \ls \frac{1-   a z^4 }{1+  a z^4 } \rs^{2} dt^2\rs \rb .
\ee
Here $a$ is related to the temperature by $a=(\pi T)^4/4$.
The scalar equation can be solved with $f(z)=1-a^2z^8$.
The solution is
\be
\phi(z)=\phi_0+\sqrt{\frac{3}{2}}\frac{c}{a} \log \frac{1+az^4 }{1-az^4},\label{phi}
\ee

While the UV boundary condition can be easily specified,
we can not specify any regular boundary condition at the horizon
unless $c=0$.
If that is the case, then such regularity requirement says that gluon condensation is 0 in deconfined phase.

To understand the regularity requirement at the boundary and also for the later comparison, we calculate the Lagrangian:
\be
{\cal L} =   {\cal R} + \frac{12}{R^2} - \half \pa_M \ph \pa^M \ph . \label{action}
\ee
The matter part is
\be
-\half g^{zz}(\pa_z\phi)^2=-{48c^2z^8\over (1-a^2z^8)^2},
\ee
which is singular at the IR boundary $z=z_T=a^{-1/4}$.
The gravity part is free from IR singularity:
\be
{\cal R}+\frac{12}{R^2}=-8.
\ee
Therefore if $c$ is non-zero, there is no way to cancel out the IR divergence, confirming above requirement.

\begin{figure}[!ht]
\begin{center}
\subfigure[] {\includegraphics[angle=0,
width=0.4\textwidth]{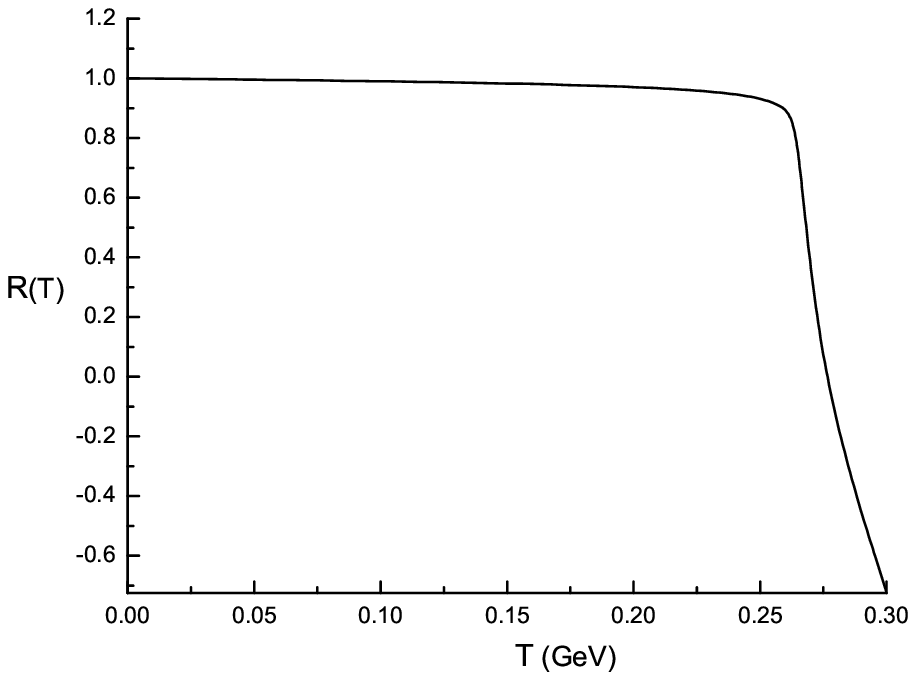} \label{fig:gCLattice}}
\subfigure[] {\includegraphics[angle=0,
width=0.4\textwidth]{{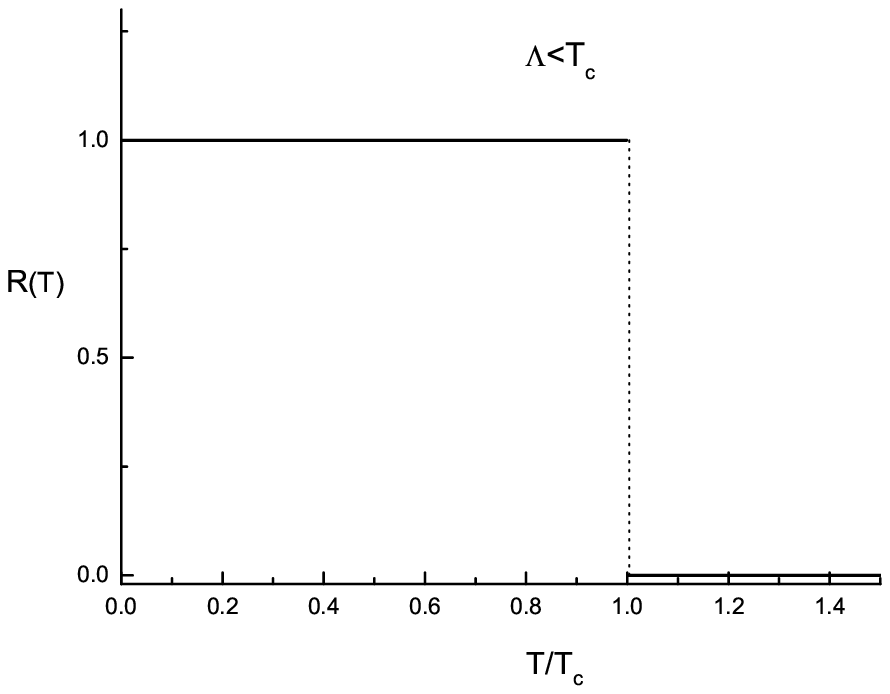}} \label{fig:gCHP2}}
 \caption{ (a) Lattice result of Miller \cite{Miller} for pure Yang Mills case, (b) The gluon condensate at finite temperature in Hard wall model. \label{gC1} }
\end{center}
\end{figure}

Therefore hard wall model prediction for the gluon condensation is
the jump from a constant finite value to 0 at the critical temperature. See the figure \ref{fig:gCHP2}.
The result is good qualitatively but {\it NOT very satisfactory}, since the lattice data shows that gluon condensation  is non-zero at high temperature.

As we will see in the next section, including the
gravity back-reaction changes the situation.

\section{Metric back-reaction  to the gluon condensation}

Now we  consider the back-reaction of the
metric to the gluon condensation.
We are mainly interested in the high temperature phase behavior of the gluon condensation.
The 5D gravity action with a dilaton is given by
\be
S = \gamma \frac{1}{2 \k^2} \int d^5 x \sqrt{g} \lb {\cal R} + \frac{12}{R^2} - \half \pa_M \ph \pa^M \ph \rb ,\label{action2}
\ee
where $\gamma=+1$ for Minkowski metric, and $\gamma=-1$  for
Euclidean signature. We work with
Minkowski metric for most cases in this paper.

The supersymmetric solution of this system is discovered in last decade
\cite{sfet,gubser3} to discuss the running coupling and confinement
and rediscovered recently in \cite{CR} to discuss the gluon condensation at zero temperature.
We call it as the dilaton-wall  solution  and follow the notation of Csaki and Reese \cite{CR} closely, where the metric  is given by
\bea
ds^2=\left ( \frac{R}{z} \right )^2  \left (
\sqrt{1-c^2z^8} ~\eta_{\mu\nu}dx^\mu dx^\nu +dz^2
\right )\, ,
\eea
and the corresponding dilaton profile is given by
\bea
\phi (z) =\sqrt{\frac{3}{2}}\log \left
  (\frac{1+cz^4}{1-cz^4}
\right ) +\phi_0\, ,
\eea
where $\phi_0$ is a constant.

Below we give a two-parameter non-BPS solution that contains both the dilaton-wall solution as well as AdS black hole solution as its limiting cases. \footnote{
After the first version of this  paper was uploaded,
we were informed that our solution also was found earlier in \cite{nojiri2,bak}
in other context. So we do not claim any originality in discovering this metric.}
We start from an ansatz
\bea
ds^2 &=&   \frac{R^2}{z^2} dz^2 + e^{2A(z)} (d \vec{x}^2 + e^{2B(z)} dt^2 ), \nn
\ph &=& \ph(z)\, .
\eea
The action and the metric ansatz lead us to the equations of motion:
\bea
-z^2 {\ph^{\pr}}^2 &=& 12 z^2 A^{\pr \pr} + 12 z A^{\pr} + 24 z^2 {A^{\pr}}^2 -24 \nn
0 &=& 4 z B^{\pr} + 4 z^2 {B^{\pr}}^2 + 4 z^2 B^{\pr \pr} + 16 z^2 A^{\pr} B^{\pr} \nn
z^2 {\ph^{\pr}}^2 &=& 24 z^2 {A^{\pr}}^2 + 12 z^2 A^{\pr} B^{\pr} -24 \nn
0 &=& z^2 \ph^{\pr \pr} + z \ph^{\pr} + 4 z^2 A^{\pr} \ph^{\pr} + z^2 B^{\pr} \ph^{\pr} ,
\eea
where a prime represents a derivative with respect to $z$.
We are looking for a solution which is asymptotically AdS
and reduces to the AdS black hole in one limit and
also leads to the above BPS solution in some other limit.

After some efforts, we find  a
  solution that is given by
\bea    \label{esol}
A(z) &=& - \log \frac{z}{R}  +
    \frac{a}{4f}  \log{ \ls \frac{1 + f  {z^4} }{1 - f  {z^4} } \rs} +
    \frac{1}{4}  \log{( 1 - f^2  {z^8} )} \nn
B(z) &=&  -  \frac{a}{f} \log{ \ls \frac{1 + f  {z^4} }{1 - f {z^4} } \rs} \nn
\ph(z) &=& \ph_0 +
    \frac{c}{f}\sqrt{\frac{3}{2}} \log{ \ls \frac{1 + f  {z^4} }{1 - f  {z^4} } \rs}.
\eea
As a result, the dilaton black hole metric reads
\be \label{dbh2}
ds^2 = \frac{R^2}{z^2} \lb dz^2 +
\ls 1-  {f^2 z^8} \rs^{1/2}  \ls \frac{1+   f z^4  }{1-  {f z^4} } \rs^{a/2f}
   \ls d \vec{x}^2 -
    \ls \frac{1-   f z^4 }{1+  f z^4 } \rs^{2a/f} dt^2\rs \rb .
\ee
Here, $f$, which determines the position of the singularity,
  is related to  $a$ and $c$ by a Pythagorean relation:
\be
f^2={a^2+ c^2}.
\ee
The parameters $a$ and $c$,  the $z^4$ coefficient of the metric and dilaton field respectively, determine the temperature and the gluon condensation
 respectively when the other is 0.
Notice that the solution is well defined only in the
range $0 < z <  f^{-1/4}:=z_f $.
Since $1/z$ is the energy scale of the
boundary theory in AdS/CFT,  $z_{f}$ can be considered as
the IR cut-off.
One should also notice that
\begin{itemize}
  \item For $a = 0$,   the solution reduces to  dilaton-wall solution.
  \item For $c = 0$,   it becomes the Schwarzschild black hole solution (\ref{dbh}) \ct{jns}.
 In this case, the parameter $a$ and the temperature is related by $a={1\over4}(\pi T)^4  \label{fc}$.
\end{itemize}
Therefore we expect that our solution describes the
finite temperature with the gluon condensation.
However,
 for the generic value of $f$, the metric has an essential singularity at $z=f^{-1/4}$ and  the Hawking temperature can NOT be determined by requiring the absence of conical singularity at the
horizon.  See the appendix. 
So we have to answer following  question  associated with the metric:
\begin{itemize}
  \item How temperature can exist in the presence of essential singularity.
\end{itemize}
In field theory, there is no barrier to define temperature
 in the presence of the gluon condensation. Therefore
geometric  indeterminacy should not mean that we do not  have   temperature in the presence of condensation.
We speculate  that 
the thermalization is incomplete unless the the gluon condensation disappear. 
This is deeply related to the nature of our solution where two independent scales (temperature and gluon condensation) co-exist. 
Before we answer this question we have  even more urgent question.
When  $c$ is non-zero, 
   the dilaton  blows up at the horizon as we already have seen in eq. (\ref{phi}). 
The dilaton blows up at $z_f$ here also.
Therefore we should ask:
\begin{itemize}
       \item Why gluon condensation can exist in spite of the
       IR divergence of the dilaton? 
\end{itemize}
To answer this question we evaluate the graviton and dilaton action. The matter part is
\be
{\cal L}_{dilaton}=-\half g^{zz}(\pa_z\phi)^2=-{48c^2z^8\over (1-f^2z^8)^2},
\ee
which is of course singular.  However the gravity part is also singular in this case
\be
{\cal L}_{grav} = {\cal R} + \frac{12}{R^2}= -8 + {48(f^2-a^2)z^8\over (1-f^2z^8)^2}.   \label{action3}
\ee
The point here is that the singularity in the  dilaton part is canceled by the singularity  developed in the gravity-back-reaction to the gluon condensation through  the Phytagorean relation $f^2=a^2+c^2$.
Since the total ${\cal L}$  has no IR singularity for  any finite values of $c$,  there is no reason why we should not allow it.

The allowance of the non-trivial matter field
through the singularity cancelation can be given an interesting interpretation in terms of the boundary condition on the fields:
When the IR boundary is a regular horizon, we have to impose a regular boundary condition there. 
This regularity forbids a non-trivial configuration of matter field $\phi$.
In the presence of the singularity, 
classical gravity near singularity is not valid. Therefore the field configuration near the singularity
is not reliable. 
Therefore we have to set an IR cutoff before we reach $z_f$  and
the boundary condition should be imposed there. This makes everything regular.
In this way, developing singularity is a mechanism by which field configuration can develop a non-trivial value of condensation. 
In other words, to allow a forbidden quantity by the regularity of the horizon, the system must develop an IR cutoff
outside the original horizon. This is the meaning of the singularity located outside original horizon ( $f>a$).
In fact this is why gluon condensation is allowed in the dilaton-wall solution at zero temperature. 
We believe that this is a general phenomena of gravity-matter interaction in AdS/CFT.

\vskip 1cm
Now we turn to the  question on  the temperature.
Our  answer is that even in the presence of $c$, we still identify $a=(\pi T)^4/4$. 
We support this choice  by following observation:  the temperature in the gravity theory is a parameter that fixes the scale of the geometry and
 the first non trivial coefficient of in the expansion of the metric,  $g^{(4)}_{\mu\nu}$,  should determine the scale of the metric. A surprising fact is that even in the presence of  $c$,
 $g^{(4)}_{\mu\nu}$ does not change:
 \bea
g_{0 0} = - 1 + 3 a z^4  + {\cal O}(z^8), \quad\quad
g_{i i} = 1 +   a z^4   + {\cal O}(z^{12}).\label{g4}
\eea
The  gluon condensation  changes the metric only from the  $g^{(8)}_{\mu\nu}$. 
This can be understood: For the gluons
participating in the thermal excitations, they should follow the usual  Stephan-Boltzmann law $\rho\sim T^4$, since they are just massless excitations.
Therefore, it is natural to identify  $g^{(4)}_{\mu\nu}$ as the gluon contribution to the energy momentum tensor $T_{\mu\nu}$ leaving aside the contribution of the glueball or the gluon condensation.
\be
g^{(4)}_{\mu\nu}= a(3,1,1,1)=\frac{\kappa^2}{2}{\rm diag}(\rho,p,p,p)_{gluon}.
\ee
This observation supports the identification $a=(\pi T)^4/4$,
since the thermal gluons as massless particles should have Stephan-Boltzmann law.
However,  the identification of the {\it total} energy momentum tensor with the $g^{(4)}_{i i}$ should be broken
to avoid following contradiction:  (\ref{g4}) indicates that the trace of boundary $T_{\mu\nu}$ is still zero, while we know that the scale invariance is dynamically broken by developing the condensation,  introducing one more scale  $c=(\pi \Lambda)^4$.
Since $a$ and $c$ are independent constants, it is natural to assume that the temperature is independent of the condensation and  vice versa.
While it is natural to have the notion of in field theoretic temperature in the presence of the gluon condensation,
its dual gravity develops an essential singulariy and forbid the geometric notion of temperature.
  The coexsitence of two scales and the coexistence of thermal gluon and condensed gluon
   means that themalization is incomplete and this, we believe,  is  the meaning of the singularity in the gravity description.

\section{Thermodynamics with Gluon condensation }
We now discuss the thermodynamic properties of the system. We begin by calculating the total action of gravity and the dilaton:
\bea
\frac{S_E^{\rm tot}}{{\beta}V}=\frac{4}{\kappa^2}
\int_\epsilon^{z_f}dz\frac{1}{z^5}
\left ( 1-f^2z^8\right) =-\frac{2}{\kappa^2}f+\frac{1}{\kappa^2\epsilon^4},\label{SdBH2}
\eea
where $V$ is the volume of $R^3$.
 We renormalize the total action   by subtracting the action value for the pure AdS, which is  nothing but the last divergent term in eq.(\ref{SdBH2}). To make sure, one can check that
this prescription works for the known (AdS black hole) case:
It is easy to show that in this case,
\be F/V=-\sigma  T^4,\;\; E/V=3\sigma T^3,\;\; p=\sigma T^4,\;\;\ee with $\sigma=\frac{\pi^4}{2\kappa^2}=\frac{\pi^2 N^2}{8}$, so that we can recover well-known Stephan-Boltzmann law with no condensation case $\rho=\frac{3}{8}\pi^2N^2T^4 .$

In our case,
\be
F/V=- \alpha f(T)  ,\;\; \rho=E/V=\alpha(-f+Tf'(T)), \;\;   p=\alpha f,\ee with $\alpha=\frac{2}{\kappa^2}=\frac{N^2}{2\pi^2}$.
With the  parametrization $c=  \frac{1}{4}(\pi \Lambda)^4$ and the temperature identification
\be
a=\frac{1}{4}(\pi T)^4
\ee
made in last section, the energy density is given by
\be
\rho(T,\Lambda)= \alpha ({3a^2-c^2})/{f}
= \frac{3\pi^2 N_c^2}{8} \frac{(T^8-\frac{1}{3}\Lambda^8)} {\sqrt{T^8+\Lambda^8}} .
\ee
The result is only for high temperature, since for low temperature
other back ground is dominating.
If we identify the dilaton (or gluon condensation) contribution of the energy momentum tensor
as the difference of total and the thermal gluon identified before:
\be
T^{GC}_{\mu\nu}= T^{total}_{\mu\nu}-T^{gluon}_{\mu\nu}=\alpha{\rm diag} (Tf'(T)-f-3a, f-a,f-a,f-a).
\ee
Notice that the gluon condensation contributes negative energy :
\be
\rho_{GC}=-\alpha ({3a(f-a)+c^2})/{f} <0, \ee
which is a reminiscent of the zero temperature result of Shifman, Vainstein and Zakharov \cite{SVZ}. In both case, the negativeness is coming from the renormalization.

At $c=0$ it goes to the usual Stephan-Boltzmann law.
The pressure can also be calculated to be
\be
p=\alpha f=\frac{\pi^2 N_c^2}{8}\sqrt{T^8+\Lambda^8}
\ee
so that the trace anomaly due to the gluon condensation is
\be
\rho-3p=-4\alpha \frac{c^2}{f} <0.
\ee
For large temperature it goes like $\sim -1/T^4 $.

In the absence of the proper horizon, the most interesting
thermodynamic quantity is the entropy.
The entropy density is
\be
s= \frac{4\alpha a}{T}\cdot \frac{a}{f}= \frac{\pi^2}{2}N_c^2T^3 \cdot \frac{T^4}{\sqrt{T^8+\Lambda^8}}   .
\ee
Notice that the entropy is decreased compared with the case with no gluon condensation by the factor $a/f$. This make sense,  since the entropy in condensed state should be less than that in thermal state.
In high temperature limit, all the thermodynamic quantities saturate
 to the case of $c=0$,  regardless of $c$.

\section{Hawking-Page Transition}
So far we have discussed high temperature regime.
Here we discuss the low temperature regime and the phase transition
by discussing the deconfinement phase transition along the line of
 \cite{witten2,herzog}.
We first evaluate the value of the action with the `thermal dilatonic AdS'
background (tdAdS), which can be obtained by the double  Wick rotation ( Wick rotation and compactifying the time) 
of the dilaton-wall solution. We take $R=1$ here
and hereafter.
Using the result for curvature of the tdAdS,
\be
{\cal R}_{\rm tdAdS}=-4\cdot({5c_0^4z^{16}-22c_0^4 z^8
  +5})/{ (1-c_0^2z^8)^2}\,
,\ee
the action  is given by
\bea
 {S_E^{\rm tdAdS}}/{V_3}=-\frac{4}{\kappa^2}\int_0^{\beta^\prime}
dt\int_\epsilon^{z_c}dz\frac{1}{z^5}
\left ( c_0^2 {z}^8 -1 \right)\, ,
\eea
where $z_c=c_0^{-1/4}$ and $V_3=\int d\vec x$ is the volume of our three  dimensional space.
The action for the dilatonic black hole (dBH) solution is given  by
\bea
\frac{S_E^{\rm dBH}}{V_3}=-\frac{4}{\kappa^2}\int_0^{\beta}
dt\int_\epsilon^{z_f}dz\frac{1}{z^5}
\left (f^2z^8 -1 \right)\, ,\label{SdBH3}
\eea
where $z_f=f^{-1/4}$. We used the fact that
the curvature in this back ground is
\be
{\cal R}_{\rm dBH} = -\frac{4\left[5 -22f^2z^8 +5f^4z^{16} +12a^2z^8 \right]}{(1-f^2z^8)^2}\,.
\ee
We determine $\beta^\prime$ of tdAdS in terms of $\beta$
by requiring  that the periods of the two backgrounds
in the compactified time direction are the same at $z=\epsilon$:
\be
\beta^\prime=\beta \exp[A(\epsilon)+B(\epsilon)-A_0(\epsilon)]
\simeq \beta(1-\frac{3}{2} a  {\epsilon^4}  )\, ,
\ee
where $A_0(\epsilon)=A(\epsilon;a=0)$.
We calculate $\Delta S:=  {S_E^{\rm dBH}}  - {S_E^{\rm
    tdAdS}}$:
\bea
\Delta S&=& \frac{2}{\kappa^2}\left [c_0
+\frac{3}{4} a -f \right ]\beta V_3.\label{dE2}
\eea
 Hawking-Page transition is at
 \be
 a_{crit}={12\over7} c_0(1+\frac{1}{3}\sqrt{16-7(c/c_0)^2}),\ee
 and condition for the real solution is  $c<1.51c_0$, but since we expect $c<c_0$,  there is always a phase transition.
 When $c<<c_0$, we get
 \be
 T_c \simeq \sqrt{2}\Lambda, \;\; with \;\; c_0=(\pi \Lambda)^4/4.
 \ee
We see that the scale of the gluon condensation provide a cut-off scale.
In hard wall model analyzed in the previous section, $1/z_m$ plays the same role.
    Due to the first order nature of the phase transition,
we could not predict the value of $c$.
The condensation is piecewise constant function of  $T$, which has a drop at a certain temperature $T_c$,
which we can estimate below.

The general structure of temperature dependence of our metric
with the Hawking-Page transition is qualitatively similar except that
at high temperature
$c=0$ for hard wall model while $c\ne 0$ with gravity back-reaction.

\section{Discussion}

In this paper we discussed the back-reaction of the metric with
the gluon condensation at the finite temperature.
We presented a dilatonic black hole solution   interpolating the
recently discovered dilaton-wall solution and AdS black hole solution.
The result contains two parameters and breaks   the conformal invariance.
We evaluated the trace anomaly and it is proportional to the gluon condensation. We showed how the system uses the singularity to allow
 the gluon condensation which is forbidden in the
presence of regular horizon.

The solution does not admit the Hawking temperature, unless the condensation parameter vanishes.
We discussed how to define the temperature in the presence of the
singularity using the holographic correspondence and an extended
 Stephan-Boltzmann law in the field theory.
 Although the AdS/CFT deals with strong coupling and the field theory with weak coupling,
 the structure is the same and we think this  structure is not modified as we change the coupling
 although the coefficient of the each term can be changed as a function of coupling.

In this paper the interpretation of singularity is such that the solution is  physical 
in the region outside a region where singularity is contained.
Our view is actually practiced extensively before in the present string community. For example in \cite{evans,evans2} Constable-Myers solution, which has a essential singularity, is used to discuss the 3+1 dimensional confining gauge theory.  However, one might wonder 
 whether one can find  singularity free solution by adding some other matter. 
 This is especially relevant because,   a singularity often means we are using wrong degree of freedom 
 or missing some degree of freedom. For example, in our case, perhaps one should treat the condened part of gluon as independent degree of freedom.  We hope we can report on this issue near future.
While we do not know any example worked out to realize the the senario suggested by referee it is definitely
interesting idea worthwhile to be pushed in the future. 

We discussed the temperature dependence of the gluon condensation with the Hawking-Page transition(HPT).
As in the usual case, there is a jump of the gluon condensation
at the critical temperature.
Apart from this jump and the
determination of the transition temperature,
there is not much feature in the temperature dependence.
Such character of a physical parameter is shared for all AdS/CFT model with HPT.
The large N nature of AdS/CFT forbids to reproduce the details of the temperature
dependence found in Lattice results.

We describe some future directions: 
 we did not include a hard wall at $z_m$ since
there is already another scale provided by the gluon condensation.
  It is interesting to discuss the Hawking-Page transition
  in the presence of three or more independent physical scales.
Also it would be interesting to discuss back-reaction of the gravity for other fields like massive scalar as well as vectors.
These topics  are under progress.
It would be also interesting to calculate various physical quantities in the presence of the gluon condensate.
We want to come back to this issue in future publications.

\appendix
\section{ Regularity of the metric  and Hawking temperature }
Here we discuss the equivalence of the three facts in the class of our metric:
(i) Definability of hawking temperature,
(ii) the regularity of the curvature,
(iii) the rationality of the metric components.
To see this, note that the relevant part of the dBH metric is
\be
ds^2=-g_{00}dt^2 +\frac{1}{z^2}dz^2\, ,
\ee
where
\be
 g_{00}=\frac{1}{z^2}(1-fz^4)^{\frac{1}{2} +\frac{3}{2}\frac{a}{f}}
 (1+fz^4)^{\frac{1}{2} -\frac{3}{2}\frac{a}{f}}\, .
\ee
Near $z_f$, behavior of the metric can be examined by introducing the coordinate $z=z_f(1-\rho^\beta)$ and rewriting the metric near $\rho\sim 0$,
\be
ds^2\simeq \beta^2 (\rho^{2\beta-2}d\rho^2
+\rho^{\alpha\beta}\frac{F(z_f)}{\beta^2}dt_E^2 )\, \label{ds2E}
\ee
where $dt_E^2=-dt^2$ and
$F(z):=g_{00}(z)/(1-\frac{z}{z_c})^{\frac{1}{2} +\frac{3}{2}\frac{a}{f}}$, which  can be written as
\be
F(z)\equiv \frac{1}{z^2}(1+z^4/z_f^4)^{\frac{1}{2} -\frac{3}{2}\frac{a}{f}}
(1+z/z_f +z^2/z_f^2+z^3/z_f^3 )^{\frac{1}{2} +\frac{3}{2}\frac{a}{f}} \, .\ee
 The standard lore to get the temperature is to request the absence of
 a conical singularity in \eq{ds2E}. However,
 this condition is met only when $\beta=1$ and $\alpha=2$, which means $f=a$. This in turn gives
$c=0$. Only for this case we have the well known result for the black hole temperature:
$ {T} = {\sqrt{2}}/{\pi z_f}$.
The Riemann curvature scalar is finite only if $f=a$ and also
$f=a$ gives the rational metric components as is manifest in the AdS Schwarzschild  solution.  One should better watch whether these equivalences are more general phenomena.

\vskip .5cm
\noindent{\bf \Large Acknowledgements}

We thank  to  M. Rho for helpful correspondences, to D. Bak and  M. Reece  for
informing us earlier literatures after uploading the first version.
SJS  wants to thank  Nuclear Physics group of SUNY at Stony Brook
for the hospitality and support during his visit, where the work is finalized.
This work was supported by the Science Research Center Program of
the Korea Science and Engineering Foundation through
the Center for Quantum Spacetime (CQUeST) of
Sogang University with grant number R11 - 2005 - 021.
The work of SJS was also partially supported by KOSEF Grant R01-2004-000-10520-0


\nc{\np}[3]{Nucl. Phys. {\bf B#1}, #2 (#3)}

\nc{\plb}[3]{Phys. Lett. {\bf B#1}, #2 (#3)}

\nc{\prl}[3]{Phys. Rev. Lett. {\bf #1}, #2 (#3)}

\nc{\prd}[3]{Phys. Rev. {\bf D#1}, #2 (#3)}

\nc{\ap}[3]{Ann. Phys. {\bf #1}, #2 (#3)}

\nc{\prep}[3]{Phys. Rep. {\bf #1}, #2 (#3)}

\nc{\ptp}[3]{Prog. Theor. Phys. {\bf #1}, #2 (#3)}

\nc{\rmp}[3]{Rev. Mod. Phys. {\bf #1}, #2 (#3)}

\nc{\cmp}[3]{Comm. Math. Phys. {\bf #1}, #2 (#3)}

\nc{\mpl}[3]{Mod. Phys. Lett. {\bf A#1}, #2 (#3)}

\nc{\cqg}[3]{Class. Quant. Grav. {\bf #1}, #2 (#3)}

\nc{\jhep}[3]{J. High Energy Phys. {\bf #1}, #2 (#3)}

\nc{\hep}[1]{{\tt hep-th/{#1}}}

\nc{\app}[3]{Ann. Phys. {\bf #1}, #2, (#3)}

\nc{\prp}[3]{Phys. Rept. {\bf #1}, #2, (#3)}

\nc{\jmp}[3]{J. Math. Phys. {\bf #1}, #2, (#3)}



\begin{thebibliography}{99}

\newcommand{\J}[4]{{ #1} {\bf #2} #4 (#3)}
\newcommand{\andJ}[3]{{\bf #1} (#2) #3}
\newcommand{\AP}{Ann.\ Phys.\ (N.Y.)}
\newcommand{\MPL}{Mod.\ Phys.\ Lett.}
\newcommand{\NP}{Nucl.\ Phys.}
\newcommand{\PL}{Phys.\ Lett.}
\newcommand{\PR}{Phys.\ Rev.}
\newcommand{\PRL}{Phys.\ Rev.\ Lett.}
\newcommand{\ATMP}{Adv.\ Theor.\ Math.\ Phys.}
\newcommand{\JHEP}{JHEP}
\newcommand{\IJMP}{Int.\ J.\ Mod.\ Phys.}
\newcommand{\JETPL}{JETP\ Lett.}
\newcommand{\SJNP}{Sov.\ J.\ Nuc.\ Phys.}
\newcommand{\PTP}{Prog.\ Theor.\ Phys.}


\bibitem{AdS/CFT}
J. M. Maldacena,
{ Adv. Theor. Math. Phys.} {\bf 2} 231 (1998),
hep-th/9711200.

 S. S. Gubser, I. R. Klebanov and A. M. Polyakov,
 ``Gauge Theory Correlators from Non-Critical String Theory,''
 \J{\PL}{B428}{1998}{105}, hep-th/9802109;

E. Witten,
``Anti De Sitter Space and Holography,''
{  Adv. Theor. Math. Phys.} {\bf 2} 253 (1998), hep-th/9802150;

O.~Aharony, S.~S. Gubser, J.~M. Maldacena, H.~Ooguri, and Y.~Oz,
``Large N  field theories, string theory and gravity,''
{Phys. Rept.} {\bf 323} 183 (2000).


\bibitem{RHIC-1}
E. V. Shuryak,
\J{\NP}{A750}{2005}{64}, hep-ph/0405066.



\bibitem{RHIC-2}
M. J. Tannenbaum,
{  Rept. Prog. Phys.} {\bf 69} 2005 (2006), nucl-ex/0603003.



\bibitem{son}
G. Policastro, D. T. Son and A. O. Starinets,
\J{\PRL}{87}{2001}{081601}, hep-th/0104066.

\bibitem{SZ}
S.~J.~Sin and I.~Zahed, 
\J{\PL}{B608}{2005}{265}, hep-th/0407215;

E. Shuryak, S-J. Sin and I. Zahed,
``A Gravity Dual of RHIC Collisions,''
J.Korean Phys. Soc. {\bf 50}, 384 (2007), hep-th/0511199.

\bibitem{Nastase}
H. Nastase,
``The RHIC fireball as a dual black hole,''
hep-th/0501068.

\bibitem{yaffe}
  C.~P.~Herzog, A.~Karch, P.~Kovtun, C.~Kozcaz and L.~G.~Yaffe,
JHEP {\bf 0607} 013 (2006),
hep-th/0605158.

\bibitem{jns}
  R.~A.~Janik and R.~Peschanski,
  Phys.\ Rev.\ D {\bf 73} 045013 (2006), hep-th/0512162;

  S.~Nakamura and S.~J.~Sin,
  JHEP {\bf 0609} 020 (2006), :hep-th/0607123;

  S.~J.~Sin, S.~Nakamura and S.~P.~Kim,
  JHEP {\bf 0612}, 075 (2006),
  hep-th/0610113.

\bibitem{gubser}
  S.~S.~Gubser,
  Phys.\ Rev.\  D {\bf 74}, 126005 (2006),
  hep-th/0605182.


\bibitem{SZ2}
  S.~J.~Sin and I.~Zahed,
  ``Ampere's law and energy loss in AdS/CFT duality,''
  Phys. Lett. {\bf B648}, 318 (2007),
  hep-ph/0606049.

\bibitem{gubser2}
  J.~J.~Friess, S.~S.~Gubser, G.~Michalogiorgakis and S.~S.~Pufu,
  ``Expanding plasmas and quasinormal modes of anti-de Sitter black holes,''
  JHEP {\bf 0704}, 080 (2007),
  hep-th/0611005.

\bibitem{AdS/QCD}
A short list of the references is:

T. Sakai and S. Sugimoto,
\J{\PTP}{113}{2005}{843}, hep-th/0412141;

J.~Erlich, E.~Katz, D.~T.~Son and M.~A.~Stephanov,
\J{\PRL}{95}{2005}{261602}, hep-ph/0501128;

L.~Da Rold and A.~Pomarol,
\J{\NP}{B721}{2005}{79}, hep-ph/0501218.

T. Hambye, B. Hassanain, J. March-Russell, M. Schvellinger, \J{\PR}{D74}{2006}{026003};
hep-ph/0612010.


\bibitem{kk}
A.~Karch and E.~Katz, 
{JHEP}{\bf 06} 043 (2002),
hep-th/0205236.


\bibitem{evans}
J.~Babington, J.~Erdmenger, N. Evans, Z.~Guralnik, and I.~Kirsch,
{Phys. Rev.} {\bf D69} 066007 (2004),
hep-th/0306018.


\bibitem{myers1}
M.~Kruczenski, D.~Mateos, R.~C. Myers, and D.~J. Winters,
{JHEP}{\bf 07} 049 (2003),
hep-th/0304032.

\bibitem{myers2}
M.~Kruczenski, D.~Mateos, R.~C. Myers, and D.~J. Winters,
{JHEP} {\bf 05} (2004) 041,
hep-th/0311270.
\bibitem{CR} C. Csaki and M. Reece, JHEP {\bf 0705}, 062 (2007)
[arXiv:hep-ph/0608266].

\bibitem{nojiri}
  S.~Nojiri and S.~D.~Odintsov,
  Phys.\ Lett.\  B {\bf 449}, 39 (1999)
  [arXiv:hep-th/9812017];\\
  S.~Nojiri and S.~D.~Odintsov,
  Phys.\ Lett.\  B {\bf 458}, 226 (1999)
  [arXiv:hep-th/9904036].

\bibitem{sfet}  A. Kehagias, K. Sfetsos,  Phys.Lett. {\bf B454},  270 (1999), hep-th/9902125.

\bibitem{gubser3}  S. S. Gubser, hep-th/9902155.

\bibitem{SVZ} M. A. Shifman, A.I. Vainshtein and V. I. Zakharov,
  Nucl. Phys. {\bf B147}, 385, 448 (1979).

\bibitem{SHLee} S. H. Lee, Phys. Rev. {\bf D40}, 2484
  (1989).

\bibitem{EGM} M. D'Elia and A. Di Giacomo and E. Meggiolaro,
  Phys. Rev. {\bf D 67}, 114504 (2003).

\bibitem{Miller}
D. E. Miller, Acta Phys.Polon. {\bf B28}, 2937 (1997);
D. E. Miller, ``Lattice QCD Calculation for the Physical Equation of
State,'' hep-ph/0608234.

\bibitem{BR06} G. E. Brown, J. W. Holt, C.-H. Lee and M. Rho,
 `` Late Hadronization and Matter Formed at RHIC: Vector
 Manifestation, Brown-Rho Scaling and Hadronic Freedom,''
 Phys. Rept. {\bf 439}, 161 (2007),
nucl-th/0608023 .

\bibitem{hawkingpage}
  S.~W.~Hawking and D.~N.~Page,
  Commun.\ Math.\ Phys.\  {\bf 87}, 577 (1983).

\bibitem{witten2}
  E.~Witten,
  Adv.\ Theor.\ Math.\ Phys.\  {\bf 2}, 505 (1998),
  hep-th/9803131.


\bibitem{herzog}
  C.~P.~Herzog,
  ``A holographic prediction of the deconfinement temperature,''
  arXiv:hep-th/0608151.
\bibitem{PS}
  J.~Polchinski and M.~J.~Strassler,
  Phys.\ Rev.\ Lett.\  {\bf 88}, 031601 (2002)
  [arXiv:hep-th/0109174].

\bibitem{neri}
  F.~Neri and A.~Gocksch,
  Phys.\ Rev.\  D {\bf 28}, 3147 (1983).
\bibitem{pisarski}
  R.~D.~Pisarski,
  Phys.\ Rev.\  D {\bf 29}, 1222 (1984).
\bibitem{KSJL}
  Y.~Kim, S.~J.~Sin, K.~H.~Jo and H.~K.~Lee,
  arXiv:hep-ph/0609008.


\bibitem{yaffe2}
  P.~Kovtun, M.~Unsal and L.~G.~Yaffe,
  ``Volume independence in large N(c) QCD-like gauge theories,''
  arXiv:hep-th/0702021.

\bibitem{nojiri2}
 S.~Nojiri and S.~D.~Odintsov,
  Phys.\ Rev.\  D {\bf 61}, 024027 (2000)
  [arXiv:hep-th/9906216].
  \bibitem{bak} D. Bak, M. Gutperle, S. Hirano, N. Ohta, Phys. Rev. {\bf D70},  086004
(2004); D. Bak, H. U. Yee, Phys.Rev. D71 , 046003 (2005).


\bibitem{skenderis}
S. de Haro, K. Skenderis and S. N. Solodukhin,
``Holographic Reconstruction of Spacetime and Renormalization
in the AdS/CFT Correspondence'',
{\sl Commun. Math. Phys.} {\bf 217} (2001) 595,
hep-th/0002230.


\bibitem{KW99} I. R. Klebanov and E. Witten, Nucl. Phys. {\bf B556},
  89 (1999).

\bibitem{rapp} H. Leutwyler, ``Deconfinement and Chiral Symmetry''
in $\underline{~QCD~~20~Years~Later~}$, Vol. 2, P. M. Zerwas and
H. A. Kastrup (Eds.), World Scientific, Singapore, 1993, pp. 693-716;
R. Rapp and J. Wambach, Adv. Nucl. Phys. {\bf 25}, 1 (2000).

\bibitem{CGC} M. Campostrini, A. Di Giacomo and Y. Gunduc,
 Phys. Lett. {\bf B225}, 393 (1989).

\bibitem{evans2}  J.~Babington, J.~Erdmenger, N.~J.~Evans, Z.~Guralnik and I.~Kirsch,
``A gravity dual of chiral symmetry breaking,''
  Fortsch.\ Phys.\  {\bf 52}, 578 (2004)  [arXiv:hep-th/0312263].
 


\end{thebibliography}
\end{document}